\documentstyle[aps,multicol,psfig]{revtex}

\renewcommand{\Re}{{\rm Re}}
\newcommand{\Am}{\AA$^{-1}$}

\newcommand{\pv}{{\bf p}}
\newcommand{\qv}{{\bf q}}
\newcommand{\Qv}{{\bf Q}}
\newcommand{\Qm}{Q}
\newcommand{\beq}{\begin{equation}}
\newcommand{\eeq}{\end{equation}}
\newcommand{\beqa}{\begin{eqnarray}}
\newcommand{\eeqa}{\end{eqnarray}}
\newcommand{\refe}[1]{(\ref{#1})}
\newcommand{\betat}{\tilde\beta}
\newcommand{\Gc}{{\cal G}}

\def\narrowtext{\begin{multicols}{2} \global\columnwidth20.5pc}
\def\widetext{\end{multicols} \global\columnwidth42.5pc}
\input{epsf.tex}

\begin{document}
\draft

\title{Theory and data analysis for excitations 
in liquid $^4$He beyond the roton minimum}

\author{F. Pistolesi\cite{FP}} 

\address{Institut Laue-Langevin,
B.P. 156, F-38042 Grenoble Cedex 9, France}

\date{\today}

\maketitle

\begin{abstract}
The hybridization of the single-excitation branch 
with the two-excitation continuum in the momentum 
region beyond the roton minimum
is reconsidered by including the effect of 
the interference term between one and two excitations.
Fits to the latest experimental data with our model
allow us to extract with improved accuracy 
the high momentum end of the $^4$He dispersion relation.
In contrast with previous results we find that the undamped 
excitations below two times the roton energy survive up to 
$Q=3.6$ \AA$^{-1}$ due to the attractive interaction between rotons.
\end{abstract}



\narrowtext

Although excitations in superfluid $^4$He 
have been widely studied
in the last decades 
(see for instance Ref.~\onlinecite{Griffin,Glyde}),
the nature of the single particle 
spectrum termination remains unclear.  
Forty years ago, Pitaevskii 
predicted different kinds of termination 
depending on the detailed form of the spectrum 
at low momenta \cite{Pitaevskii59}. 
In the case of a decay into pairs of rotons, 
Pitaevskii theory describes the avoided crossing
of the bare one-excitation branch with the 
continuum of two excitations. 
Within this picture 
the low energy pole is repelled by the continuum, so that 
the spectrum flattens out for large $Q$ towards 2$\Delta$ losing 
spectral weight ($\Delta$ is the roton energy). 
At the same time, 
a damped excitation for $\omega>2\Delta$
appears and shifts to higher energies.
Neutron scattering experiments
later suggested that the decay of excitations 
into {\em pairs of rotons} did actually take place for 
momentum $Q \gtrsim 2.6$~\AA$^{-1}$ \cite{CowleyWoods}.
The flattening of the spectrum at an energy of the order of $2\Delta$ 
being the main feature in support of Pitaevskii's picture.
Despite the good {\em qualitative} agreement between theory 
and experiment, several issues addressed by the  
theory could not be verified due to insufficient  
instrumental resolution. 
In particular, theory predicts a singular termination of the 
spectrum at a definite value of momentum $Q_c$ for 
a repulsive interaction ($V_4>0$) between rotons. 
In the case of an attractive interaction,  
hybridization between the roton bound state and the single quasiparticles 
is instead expected, 
as proposed by Zawadowski-Ruvalds-Solana (ZRS) 
\cite{RuvaldsZawadowski}, 
with the consequence that an undamped
quasiparticle peak at an energy slightly 
below 2$\Delta$ should be present
\cite{RuvaldsZawadowski,Pitaevskii70,Bedelletal,JugeGriffin}.
To our knowledge it has not yet been possible to distinguish clearly  
between these two cases.

Smith {\em et al.\/} \cite{Smithetal} performed the first complete
experimental analysis for $2.7\leq Q \leq  3.3$~\Am. 
Although they found indications of a repulsive interaction
($V_4>0$) using  ZRS theory, 
they pointed out that the experimental finding of 
energies for the quasiparticle peak 
above $2\Delta$ was not accounted for by 
the theory with reasonable values of the parameters.
As a matter of fact, the position of the low-energy peak was not 
extracted from the data with ZRS theory, 
but rather by fitting a Gaussian peak on a background
of constant slope. The resulting spectrum reached values
{\em above} $2 \Delta$, in contrast with theoretical predictions.
%
%
From the theoretical point of view 
it is not possible to explain in a simple way 
the presence of {\em sharp} peaks at energies above $2 \Delta$, 
as the corresponding excitation should be unstable 
towards decay into two rotons. 
%
%
Moreover, the experimental finding of a repulsive interaction disagrees 
with different theoretical calculations that predict 
a negative value for $V_4$\cite{Lee,Bedelletal}. 
More recent experimental investigations by F\aa k and coworkers
\cite{fak91,fak1}
concentrate on the interesting temperature dependence of 
the dynamical structure factor $S(\Qv,\omega)$ and do not address directly
the issue of the quasiparticle energy or 
the interaction potential between rotons. 
In any case they show clearly that there is a strong correlation
between the low-energy peak and the high energy continuum 
as $Q$ increases from 2.3 to 3.6 \Am \cite{fak1}, thus indicating 
that hybridization takes place. 
We further recall that this hybridization is expected as
direct consequence of Bose condensation 
as explained in details in Ref.~\onlinecite{Griffin}.

From the theoretical point of view, we recall that 
the validity of Pitaevskii and ZRS theories 
is restricted to a small region around 2$\Delta$.  
Indeed Pitaevskii in his original paper \cite{Pitaevskii59}
exploited the logarithmic divergence 
appearing in the two-roton response function
[$
	F_o({\bf Q}, \omega)
	=
	i\,\int {d\omega'\over 2\pi} \int {d^3 \Qv' \over (2\pi)^3}
	\Gc(p-p')\Gc(p')
$,
where $\Gc^{-1}(p)=\omega+i\,0^{+}-\omega({\bf Q})$,
$\omega({\bf Q})$ is the measured spectrum and
$p=(\Qv,\omega)$]
to solve {\em exactly} the many-body equations.
This elegant theory provides
explicit expressions for the Green and the 
density-density correlation functions, 
valid {\em only} in the small energy range where the 
singularity dominates. 
This fact leads to problems in data analysis when $\Qm$ grows
so that the bare excitation energy $\omega_o({\bf Q})$ 
takes values above $2\Delta$. 
In fact, the signal around $2 \Delta$ strongly decreases with $\Qm$
and spectral weight shifts to higher energies 
following $\omega_o({\bf Q})$.
To understand the correlation between the high energy part of the 
spectrum and the one-excitation contribution,
it is necessary to extend the validity of the theory
to a wider range of energies 
in order to describe properly the continuum contribution
to $S(\Qv, \omega)$. 
It then becomes crucial to consider the effect of 
the direct excitation of two quasiparticles by the neutron 
and its interplay with the one-quasiparticle 
excitations usually considered.

The aim of the present paper is to construct such an extension of
the theory to describe experimental data for 
$S(\Qv,\omega)$ at very low temperatures. 
Since excitations in $^4$He are stable
($\Gamma(\Qv)/\omega(\Qv)\approx 10^{-2}$ for rotons at $1.3$~K
with $\Gamma$ the half width of the excitation), 
it is an excellent approximation to write the Hamiltonian directly in terms 
of the creation and destruction 
operators $b^\dag$ and $b$ of these excitations:
\beq
	H_o = \sum_{\pv} \omega_o(\pv)
	b^\dag_{\pv} b^{\phantom{\dag}}_{\pv}
	+ V_3 \sum_{\pv,\qv} 
	\left[
	b^\dag_{\pv} b^{\dag}_{\qv} b^{\phantom{\dag}}_{\pv+\qv} 
	+ cc
	\right].	
	\label{eq1}
\eeq
We consider for the moment only the $V_3$ interaction that 
induces  the hybridization of the single with the double
excitation. 
%
%
%
In general this vertex will be a function of two momenta,
for instance the total momentum of the two particles and the momentum
of one of them. We neglect from the outset this dependence on momentum
as it is expected to be smooth in the region of interest and less 
important than the frequency dependence 
retained in the following.
%
%
Since $S(\Qv,\omega)$ is the imaginary part of the density-density
response function [$\chi(\Qv,\omega)$] 
it is convenient to express the density operator 
$\rho_{\pv}$ in terms of the $b$ and $b^{\dag}$ field operators.
In general this will be an infinite series in the $b$-fields, 
by retaining only the one and two quasi-particle 
terms we obtain:
\beqa
	\rho_{\bf p} 
	=
	\alpha(\pv) \left[b^\dag_{\pv}+b^{\phantom{\dag}}_{-\pv}\right] 
	+ 
	\sum_{\qv}
	\gamma(\pv,\qv)\,b^{\dag}_{\pv+\qv} b^{\phantom{\dag}}_{\qv}
	\nonumber\\
	+\sum_{\qv}
	\beta(\pv,\qv)\left[ 
	b^\dag_{\pv+\qv} b^\dag_{-\qv} 
	+
	b^{\phantom{\dag}}_{\qv} b^{\phantom{\dag}}_{-\pv-\qv}
	\right].
	\label{rho}
\eeqa
Eq.~\refe{rho} gives the most general second order form for $\rho_{\bf p}$ 
in terms of $b$ and $b^\dag$ that fulfills parity, 
time-reversal, and $\rho_\pv^\dag=\rho_{-\pv}$ transformation properties. 
For the same invariances the three functions introduced $\alpha$, 
$\beta$, and $\gamma$ are bound to be real. 
Although the above expression for $\rho_\pv$ 
is quite general it can be obtained microscopically within a
particular approximation scheme \cite{bogexp}.

\begin{figure}
\centerline{\psfig{file=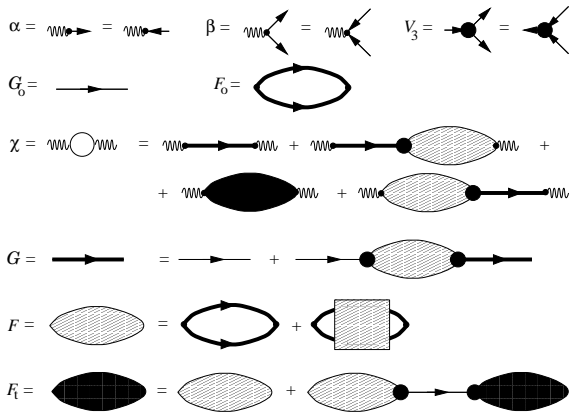,width=7.5cm,angle=-90}}
\caption{Full lines stands for 
the exact Green function, dashed boxes for the sum of 
all one-particle irreducible diagrams, and the black box for the 
sum of all diagrams with two lines closing at the two ends.
$\chi(p)$ is the sum of all diagrams with two density insertions 
(wiggly lines).}
\label{fig1}
\end{figure}

It is possible at this point to calculate perturbatively 
$\chi(\Qv,\omega)$ using the explicit form
of $\rho_\pv$ given in Eq.~\refe{rho} and the Hamiltonian \refe{eq1}. 
At zero temperature the contribution of the $\gamma(\pv,\qv)$ 
term to $\chi$ vanishes, while   
the momentum dependence of  
$\alpha(\pv)$ and $\beta(\pv,\qv)$ is neglected 
for the sake of simplicity.
The diagrammatic theory for the model is shown 
in Fig.~\ref{fig1} together with 
the definition of $\chi(p)$,
$F_o(p)$, $F_t(p)$, $F(p)$, and $\Gc_o(p)=[\omega-\omega_o(\Qv)]^{-1}$. 
In this figure dashed boxes represent the sum of all 
one-particle irreducible diagrams. 
$F_t$ stands for the sum of all diagrams 
with two lines closing at the two ends, and 
$F$ is the self-energy without the two 
external couplings.
Since the momentum dependence of the vertices is neglected, the Dyson 
equations for $\Gc$ and $F_t$ (see Fig.~\ref{fig1}) 
take the simple form of a coupled algebraic system:
\begin{mathletters}
\label{dyson}
\beqa
	\Gc(p) &=& \Gc_o(p) + \Gc_o(p)\, V_3\, F(p)\, V_3\, \Gc(p),\\
	F_t(p) &=& F(p) + F(p)\,V_3\,\Gc_o(p)\,V_3\,F_t(p).
\eeqa
\end{mathletters}
By solving Eqs.~\refe{dyson} and substituting 
$\Gc$ and $F_t$ into the expression
for $\chi(p)$ we finally obtain:
\beq
	\chi(p) = 
	{\alpha^2 + 2 \alpha \beta V_3 F(p) 
	+\beta^2 F(p)\,\Gc_o^{-1}(p) 
	\over 
	\Gc_o^{-1}(p) - V_3^2\, F(p)
	}
	\label{eq11}.
\eeq
Eq.~\refe{eq11} is the basic expression that we will use in the 
following for the fits to the data 
[$S(\Qv,\omega)=-{\rm Im} \chi(\Qv, \omega)$].
The presence of a $V_4$ interaction does not change 
the above treatment since no $V_4$ vertex can connect 
$F$ to $\Gc$ at zero temperature. This implies 
that the additional diagrams due to a $V_4$ interaction 
contribute only to $F$. 
Since the explicit calculation of $F$ is difficult and 
in general depends on the detailed structure of the vertex functions,
we prefer to extract it directly from data by exploiting 
the large (energy and momentum) region of validity of Eq.~\refe{eq11}.
We proceed by noting that in
Eq.~\refe{eq11} only $F$ and $\omega_o$ (through $\Gc_o$) 
depend on $\Qm$. 
Explicit evaluation of $F_o$ suggests that the momentum dependence 
of $F$ should not be pronounced in the range of interest 2.3-3.2 \Am.
We thus drop this dependence completely in $F$ and fit 
the resulting function $F(\omega)$ to the data.
The $Q$-dependence of $\omega_o$ is not negligible because it
drives $\omega_o(Q)$ through the $2\Delta$ line.
In this way we are left with only one parameter dependent
on $Q$. 
It is now possible to extract both $\omega(\Qv)$ and $F(\omega)$
by fitting Eq.~\refe{eq11} to all sets of data of different momentum
{\em at the same time}. 
This procedure exploits fully the information contained in the 
data because it is sensitive to the correlation among sets with 
different $Q$. 
For this reason we are able to extract information for $\omega(\Qv)$ and 
for the function $F(\omega)$ over a scale (slightly) 
smaller than the instrumental resolution.

A few words are necessary to explain how we can find the 
complex function $F(\omega)$ from the data.
The imaginary and real part of $F$ are related by the 
relation
$
\Re F(\omega) 
= -{1/\pi}{\cal P}\int d\omega' {\rm Im} F(\omega)/(\omega-\omega')
$.
We thus need only to parametrize one of them, 
we choose ${\rm Im} F(\omega)$, since
on this function it is easier to apply 
the physical constraint that excitations
are kinematically stable for energies smaller than 2$\Delta$. 
This condition reads ${\rm Im} F(\omega)=0$ for $\omega<2\Delta$ 
(the small spectral density between 
$\omega(\Qv)$ and $2\Delta$ can be neglected).
A simple way to parametrize the function ${\rm Im} F$ is then to 
choose a set of values of $\omega$ say 
$\{\omega_1, \omega_2, \dots, \omega_N\}$ 
with $2\Delta = \omega_1 < \omega_2 <\dots<\omega_N$ reasonably 
spaced and to assign a {\em free} parameter, $a_i$, to each of them.
${\rm Im} F(\omega)$ can then be defined as a cubic spline 
interpolation on such a set. 
The integral to calculate the real part 
can be performed analytically 
so that its evaluation is fast and reliable
also near the logarithmic singularity at the threshold.

%
\begin{figure}
\centerline{\psfig{file=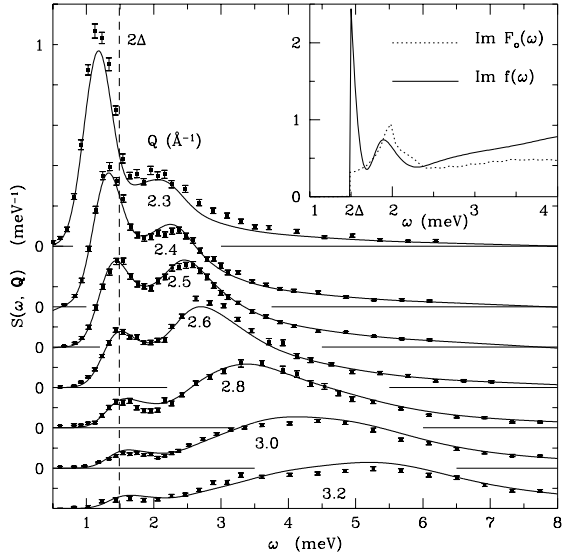,width=8.cm}}
\caption{
Fit to the data from Ref.~[12] 
with a parametrized ${\rm Im} f$. In the inset the 
resulting ${\rm Im} f$ is shown compared with ${\rm Im} F_o$ averaged 
over $2.3<Q<3.2$ \AA$^{-1}$ and properly scaled.}
\label{fig2}
\end{figure}
%

We take advantage of a symmetry in Eq.~\refe{eq11} to define 
a dimensionless function $f(\omega)=\lambda F(\omega)$,
constrained by the normalization condition 
$
	\int_{2\Delta}^{\omega_N} d\omega {\rm Im} f(\omega)
	=	
	\omega_N-2\Delta
$. We thus define $g_3=V_3/\lambda^{1/2}$, 
and $\betat=\beta /(\lambda^{1/2}\alpha)$.
In this way $\alpha$, $\betat$, $g_3$, and the $N-1$ parameters that define 
$f$ are independent and can be fitted to the data.
We set ${\rm Im} f(\omega)=a_N$ for $\omega>\omega_N$ and 
subtract out the infinite constant that 
would appear in $\Re F(\omega)$.

We thus fitted Eq.~\refe{eq11} 
(convolved with the known instrumental resolution) 
to experimental data of Ref.~\onlinecite{fak1} (at 1.30 K) by 
minimizing numerically $\chi^2$ for the 
parameters: 
$\alpha$, $\betat$, $g_3$, $\omega(\Qv_1)$, $\dots$, $\omega(\Qv_n)$,
$a_1$, $\dots$, $a_{N-1}$ ($N=15$ for the fit presented). 
The minimization procedure has been performed with 
different standard routines and the result turned out 
to be independent of the choice of $\{\omega_i\}$ and 
the initial value of $a_i$.
A typical starting point for ${\rm Im} f$ is simply 
$a_i=1$ for all $i$.
The resulting fit is shown in Fig.~\ref{fig2}. 
Good agreement 
between theory and experiment is obtained with a  
reduced $\chi^2$ ($\chi^2_R$) of nearly 4,
thus indicating that even if we are leaving a large freedom in 
$f$ the agreement is significant.

\begin{figure}
\centerline{\psfig{file=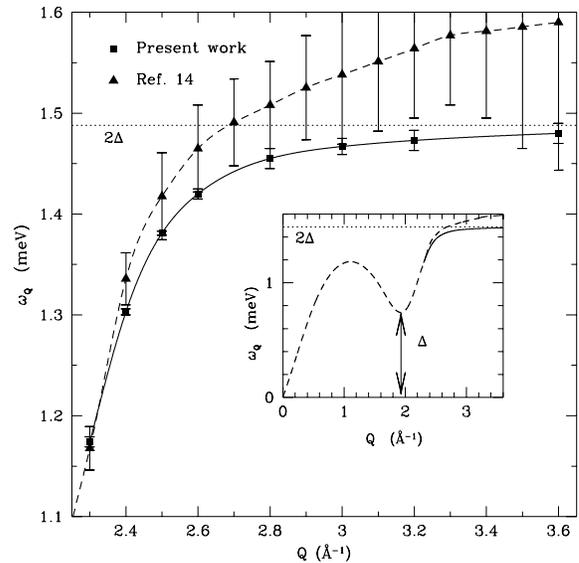,width=8.cm}}
\caption{Dispersion relation found in the present work (solid line),
and in Ref.~[4,14] (dashed line). The inset shows  the 
complete dispersion relation. 
}
\label{fig3}
\end{figure}

Our main result is summarized by the dispersion relation
for the undamped excitation shown in Fig.~\ref{fig3}, where 
it is compared with the one reported
in Ref.~\onlinecite{Donnellyetal}.
%
%
We find that the model can {\em quantitatively} explain 
that the peak position of $S(Q,\omega)$ is 
slightly larger than $2\Delta$ for $Q>2.6$ \AA$^{-1}$.
This originates from a mixing within the instrumental resolution
of the contribution of the sharp peak
at energy $\omega_Q$ slightly smaller than 2$\Delta$ with that 
of the continuum of two
rotons excitations starting at $2\Delta$.
On general grounds the continuum should depend 
strongly on $\omega$ near $2\Delta$, 
as for $\omega>2\Delta$ there are much more states available 
to decay into.
For these reasons, the assumption of a
background of constant slope, used to obtain the values of 
Ref.~\onlinecite{Donnellyetal}, is not valid in this 
momentum region.
Our procedure exploits instead the theoretical model for $\chi(p)$ 
to extract the value of $\omega_Q$. In this way we can  
find the final part of the dispersion relation with improved 
accuracy and it turns out that data agree with a dispersion
relation for the excitations always below 2$\Delta$.
%
%
%

The importance of the fitted parameter $\betat$ has been checked
by studying the function $\chi^2(\betat)$, 
where all the other parameters are properly modified 
to minimize $\chi^2$ for each value of $\betat$.
The confidence region for $\betat$, 
{\em i.e.} the values of $\betat$ such that  
$\chi^2(\betat)/\chi^2_{\rm min} < 1.5$,
turns out to be $-0.17<\betat<-0.01$ meV$^{-1/2}$ with a best value
$\betat=-0.06$ meV$^{-1/2}$. This implies that the direct 
excitations of two quasiparticles by the neutron 
gives a small but {\em sizable} contribution to $S(\Qv,\omega)$.
Concerning the other parameters 
we find $\alpha^2=1.4$ and  $g_3=0.8$ meV$^{1/2}$.

The shape of ${\rm Im} f$ found by the fit has two main 
features: a clear peak at $\omega\approx 2$ meV and 
a ``quasi-divergent'' behavior at the threshold. 
The peak is due to the maxon-roton van Hove 
singularity. 
This can be verified by comparing the function
${\rm Im} F_o(Q, \omega)$, averaged over 
$2.3 \leq Q  \leq 3.2$ \Am 
and properly scaled, with the fitted ${\rm Im} f(\omega)$ 
(see the inset of Fig.~\ref{fig2}).
Hence it is clear that the peak corresponds in shape and position 
to the maxon-roton singularity. 
It is remarkable that no trace of the 
peak is apparent in any of the experimental plots.
It is only by exploiting the correlation between 
plots with different $Q$ that we have been able to extract 
this information.
On this basis, using the fitted parameters 
it is also possible to predict that 
a peak and its shape should be observable  
with a resolution of $0.1$ meV
(to be compared with $0.5$ meV of Fig.~\ref{fig2}). 

The quasi-singularity at threshold can  
be understood as an interaction effect,
namely a signature of the attractive interaction between rotons.
As a matter of fact, in the 
small region of energy near the threshold we can 
apply Pitaevski-ZRS 
\cite{Pitaevskii59,RuvaldsZawadowski,Pitaevskii70}
theory to evaluate $F(\omega)$ that reads:
\beq
	F(\Qv, \omega) = 
	4 \rho\,\ln {2\Delta-\omega \over D}
	\left(1-g_4 \ln{2\Delta-\omega \over D}\right)^{-1},
	\label{eq5}
\eeq
where $\rho=-{\rm Im} F_o(\Qv,\omega=2\Delta^+)/2\pi$ is the 
threshold density of states, 
$g_4=2 V_4 \rho$, and $D$ is a cutoff that can be set to 1 meV as 
changes in $D$ can be easily reabsorbed in small variations of 
$\omega_o$, $g_3$, and $g_4$. 
One can thus readily verify that with a negative value of $g_4$ 
Eq.~\refe{eq5} gives a ${\rm Im} F$ 
that for $\omega \rightarrow 2\Delta^+$ reproduces 
qualitatively ${\rm Im} f$ of Fig.~\ref{fig2}.
To verify quantitatively this fact and to find an 
estimate of $g_4$ we have repeated the
data fit using Eq.~\refe{eq5} to parametrize $f(\omega)$, 
(instead of the spline parametrization) 
and setting a cutoff in energy at $2\Delta$+0.2 meV, 
in such a way to apply Eq.~\refe{eq5} {\em only} where it is 
supposed to hold.
The resulting $\chi^2_R=1.7$ for the fit gives evidence that 
the theory works {\em quantitatively} in this region and  
we get for the interaction parameter 
$V_4 = g_4/(2\rho) \approx -4.7$ meV \AA$^3$
($g_4=-0.15$). 
The bound-state energy that we obtain 
from Eq.~\refe{eq5} is indeed very small 
$E_B=D\,\exp\{1/g_4\}\approx 1.3 \mu eV$. 
This second fit gives also an additional estimate of the dispersion
relation that agrees with the previous one and confirms   
that an undamped state is present up to $Q=3.6$ \AA$^{-1}$. 
The confidence region  for $g_4$ is $-0.22<g_4<-0.08$, 
restricting $\betat$ to its confidence region already found, which 
indicates that the interaction is definitely {\em attractive}. 

Some final comments are in order.
First, 
while Fig.~\ref{fig2} shows the fit for $2.3\leq Q\leq 3.2$,
the inclusion of the additional set of data for $Q=3.6$ 
increases slightly $\chi_R^2$ (since the momentum range over 
which we are assuming the vertices to be momentum-independent
may be too large) but it remains in any case a good fit to data.
 For this reason we report in Fig.~\ref{fig3} the value for 
$\omega(\Qv)$ obtained in this way.
Second, 
we studied the cutoff dependence of $\chi^2_R$
when $f(\omega)$ is parametrized according to Eq.~\refe{eq5} 
and we found that it is very weak up to 
$\omega=2$ meV where the effect of the maxon-roton 
peak becomes important. 
Thus use of the Pitaevskii-ZRS theory for $F$ in our expression \refe{eq11}
does not give a good description of the 
data if the cutoff in energy is removed. 
This indicates that the maxon-roton structure and, in general, 
the whole shape of ${\rm Im} f$ plays a {\em crucial} 
role in determining $S(\Qv, \omega)$.

In conclusion, we presented a theory for $S(\Qv, \omega)$ that takes into 
account  both one- and two-quasiparticle excitations by the neutron.
The theory reproduces the experimental result over a 
large range of energy and momentum. 
We have thus been able to extract 
the final part of the spectrum dispersion relation in $^4$He.
%
%
Our theory can be regarded as an extension of the 
Pitaevskii-ZRS theory taking into account the effect of 
two-particle excitations. 
Moreover the range of validity is enlarged as we make no
hypothesis on the $F$ function, but we extract it directly
from data. 
In the region where Pitaevskii-ZRS theory holds we have used it 
to parametrize $F$ in Eq.~\refe{eq11} and we found that 
the interaction potential among rotons ($V_4$) is attractive 
in this momentum region.

I am indebted to P.~Nozi\`eres for suggesting this problem to me and
for many discussions. I gratefully acknowledge B.~F\aa k for  
discussions and critical reading of the manuscript. 
I acknowledge B.~F\aa k and J.~Bossy for 
letting me use their data prior to publication.
I also thank N.~Cooper, N.~Manini, A.~W\"urger, P.~Pieri, 
G.C.~Strinati, and H.R.~Glyde for useful discussions.

\widetext
\end{document}